\documentclass[12pt]{article}
\usepackage{latexsym,epsfig,amssymb,euscript,amsmath,verbatim,color}

\def\be{\begin{equation}}
\def\ee{\end{equation}}
\def\bea{\begin{eqnarray}}
\def\eea{\end{eqnarray}}
\newcommand{\nn}{\nonumber}

\begin{document}

\pagestyle{empty}

\begin{center}

{\LARGE{\bf Chiral Extensions of the MSSM}}

\vspace{1.8cm}

{\large{Gabriele Ferretti and Denis Karateev\footnote{Current address: Department of Astronomy and Theoretical Physics,
Lund University, S\"olvegatan 14A, SE-223 62 Lund, Sweden}}}

\bigskip

{\it Department of Fundamental Physics \\
Chalmers University of Technology, 412 96 G\"oteborg, Sweden }

\vspace{2cm}

{\bf Abstract}

\vspace{1.5cm}

\begin{minipage}[h]{16.0cm}
We present a class of extensions of the MSSM characterized by a fully chiral field content (no $\mu$-terms) and no baryon or lepton number violating term in the superpotential due to an extra $U'(1)$ gauge symmetry.
The minimal model consist of the usual matter sector with family dependent $U'(1)$ charges, six Higgs weak doublets, and three singlets required to give masses to the Higgsinos and cancel anomalies. We discuss its main features such as the tree level mass spectrum and the constraints on flavor changing processes.
\end{minipage}

\end{center}

\newpage

\setcounter{page}{1} \pagestyle{plain} \renewcommand{\thefootnote}{\arabic{footnote}} \setcounter{footnote}{0}

\section{Introduction}

Supersymmetry (SUSY) is a deep theoretical idea whose relevance to Electro-Weak (EW) symmetry breaking is now being tested at the LHC.
In this context, most of the attention has been focused on the minimal extension of the Standard Model (MSSM) consisting of a supermultiplet for each known matter field and gauge boson and a pair of Higgs doublets (for a review see~\cite{Martin:1997ns}). Minimality is certainly appealing for many reasons. To begin with, if SUSY is realized in nature, all of the particles of the MSSM are guaranteed to exist. Furthermore, minimality allows to control, at least to some extent, the vast parameter space coming when SUSY is spontaneously broken and to set up benchmark points to be used as guidelines in the search for new physics.

In the case of the MSSM however, minimality comes at a price. The accidental symmetries that follow from gauge invariance and chirality in the Standard Model and prevent e.g. proton decay no longer arise automatically in the MSSM and need to be enforced by imposing R-parity or other discrete symmetries. Furthermore, one of the main reasons to believe that SUSY is relevant to physics at the TeV scale is the fact that it solves the ``hierarchy problem" arising in the presence of light fundamental scalars such as the Higgs. In the MSSM however, this solution is only partial, since one is required to introduce the $\mu$-term in the superpotential, which, while technically natural (the superpotential in not renormalized) begs the question of why it should be at the same scale as EW breaking. Lastly, in view of the discovery of a Higgs boson at $126$~GeV~\cite{:2012gk,:2012gu} and the failure to observe colored superparticles, the pure MSSM itself is starting to look quite fine-tuned (see e.g.~\cite{Lodone:2012kp,Cao:2012yn} for recent reviews).

The $\mu$ problem is due to the fact that the MSSM matter content is not ``fully chiral" since the two Higgs doublets $H_u$ and $H_d$ together form a real representation of the gauge group. Chirality has certainly served us well in the Standard Model, preventing bare fermion masses and unwanted couplings but it is not fully exploited in the MSSM. Some of the popular extensions of the MSSM, such as the NMSSM (reviewed in \cite{Ellwanger:2009dp}), solve this problem at the cost of adding an ad hoc discrete symmetry, and one wonders what the minimal fully chiral model with only accidental global symmetries might look like. This question needs to be made more specific in order to be properly analyzed, so in this paper we address the following issue: what is the minimal extension of the gauge group and the Higgs sector of the MSSM for which all SUSY masses ($\mu$-terms) and baryon number (B) and lepton number (L) violating terms are forbidden?
We shall see that it is possible to satisfy these requirements by extending the gauge group by a single $U(1)'$ and the Higgs sector by a total of six Higgs doublets (instead of the two of the MSSM) and three $U(1)'$ charged singlets.

We shall work in a strictly ``bottom-up" approach and give up manifest grand-unification. Our approach necessarily leads to some amount of tree level Flavor Violation (FV) which is notoriously strongly constrained by many experimental observations. We will discuss this point in section~3 together with the tree level spectrum for the minimal model, but one interesting point worth mentioning already is that only the weak singlets are allowed to have family dependent charges, thus evading the strongest constraints on FV. (See e.g.~\cite{ArkaniHamed:1999yy,Langacker:2008yv}.)

More models can be obtained by changing the structure of the superpotential and increasing the number of singlets but the minimal model we present is singled out amongst all those found by an extensive computerized search by having a unique set of charges.

In section~2 we discuss the basic assumptions that go into the formulation of these class of models and present the minimal solution. In section~3 we discuss some of the basic phenomenological features of the minimal model.

The MSSM with extended gauge group by an extra $U'(1)$ is often denoted by UMSSM~\cite{Erler:2002a0,Langacker:1998tc,Cvetic:1997ky}. The literature on the subject is vast. Earlier work on flavor $U'(1)$ includes~\cite{Dudas:1995yu, Nir:1995bu, Savoy:2010sj}. Previous versions of chiral UMSSM (different from ours) are given in e.g. \cite{Cheng:1998a1, Lee:2008a2, Erler:2000a3, Ma:2002a4, Aoki:1999tv, Aoki:2000a5, Demir:2005a6, Everett:2009cn,Perez:2011sr}. In almost all these models it is required to include additional colored superfields (exotics) for anomaly cancelation, which we avoid.

\section{Construction of the model}

Our objective is that of constructing a fully chiral renormalizable model by extending the Higgs sector while keeping the same matter sector as the MSSM. Clearly this cannot be reached with the standard gauge content. Requiring the gauge invariance of the Yukawa couplings forces the Higgs fields of $u$ and $d$ type to transform as weak doublets with opposite hypercharge. The minimal extension is that of adding just one extra $U(1)$ gauge group, henceforth denoted by $U(1)'$, under which the fields in the Higgs sector are charged in order to prevent $\mu$-terms. Some of the matter fields in different families will also carry different $U(1)'$ charges. If this were not the case, it is easy to see that the Higgses would, once again, have equal and opposite $U(1)'$ charges allowing a $\mu$-term.

We thus assume a ``matter sector" consisting of the usual three $Q^i,u^i,d^i,L^i,e^i$ families, ($i=1,2,3$ family number) as in the MSSM. For ease of notation we refrain from putting tildes or bars on the fields--all superfields are assumed to be chiral (with left handed fermions) under SUSY and the scalar components are denoted by the same symbol~\footnote{Additional (weak singlet) neutrino superfields $v^i$ could also be added in order to construct Dirac masses.}.

The matter sector acquires a mass by renormalizable couplings to an extended ``Higgs sector", comprising of a number of Higgs pairs $(H_u^a, H_d^a)$, $a=1\dots m$ carrying the usual MSSM quantum numbers and a $U(1)'$ charge. Finally, we need a number of $U(1)'$ charged MSSM singlets $S^r$, $r=1\dots n$, to give mass to the Higgsinos via a coupling of type $H_u H_d S$ and to cancel the anomalies. We refer to these fields as ``the singlet sector".

We will impose that the extra $U(1)'$ symmetry automatically forbids any dimensionfull coupling in the superpotential $W$, namely any linear combination of the following terms
\be
    S^r,~S^r S^s,~H_u^a H_d^b,~H_u^a L^i \not\in W. \label{nomu}
\ee
Furthermore, we impose that the same gauge symmetry forbids dimension three B or L violating terms in the superpotential, i.e. any linear combination of terms like
\be
   u^i d^j d^k, ~L^i L^j e^k, ~L^i Q^j d^k, ~L^i H_u^a S^r \not\in W. \label{noBL}
\ee
Eqs. (\ref{nomu}) and  (\ref{noBL}) translate into a set of inequalities for the  $U(1)'$ charges to be satisfied by our solution.
Condition (\ref{noBL}), which essentially forbids the same (dimension four) terms as those usually called R-parity violating (RPV), might need to be relaxed in light of the recent LHC searches failing to see large amounts of missing energy. Our strategy could easily be modified along these lines.

Thus, the most general form of the superpotential in our construction is
\be
 W = y^u_{ija} Q^i u^j H_u^a + y^d_{ija} Q^i d^j H_d^a + y^e_{ija} L^i e^j H_d^a
     +\kappa_{abr} H_u^a H_d^b S^r + \lambda_{rst}S^r S^s S^t, \label{yuk}
\ee
The first three terms are required to give masses to the matter fields. We will assume that all possible such terms are actually present (no ``textures"), although even this condition could be relaxed if needed.

The presence of a new gauge group introduces a whole new set of anomaly cancelation conditions namely: $SU(3)_C^2 U(1)'$, $SU(2)_W^2 U(1)'$, $U(1)_Y^2 U(1)'$, $\mathrm{Grav}^2 U(1)'$, $U(1)_Y U(1)^{'2}$ and $U(1)^{'3}$. In the obvious notation for the charges, they read
\bea
     && \sum_{i=1}^3 \left(2 q_Q^i + q_u^i + q_d^i\right) = 0, \label{L1} \\
     && \sum_{i=1}^3 (3 q_Q^i + q_L^i) + \sum_{a=1}^m (q_{Hu}^a + q_{Hd}^a) = 0, \label{L2}\\
     && \sum_{i=1}^3 (q_Q^i + 8 q_u^i + 2 q_d^i + 3 q_L^i + 6 q_e^i) +
            \sum_{a=1}^m( 3 q_{Hu}^a + 3 q_{Hd}^a) = 0, \label{L3}\\
     && \sum_{i=1}^3  (6 q_Q^i + 3 q_u^i + 3 q_d^i + 2 q_L^i + q_e^i) +
            \sum_{a=1}^m (2 q_{Hu}^a + 2 q_{Hd}^a) + \sum_{r=1}^n q_S^r = 0,\label{L4}\\
     && \sum_{i=1}^3 (q_Q^{i 2} - 2 q_u^{i 2} + q_d^{i 2} - q_L^{i 2}+ q_e^{i 2}) +
             \sum_{a=1}^m (q_{Hu}^{a 2} - q_{Hd}^{a 2}) = 0, \label{N1} \\
     && \sum_{i=1}^3 (6 q_Q^{i 3} + 3 q_u^{i 3} + 3 q_d^{i 3} + 2 q_L^{i 3} + q_e^{i 3}) +
             \sum_{a=1}^m (2 q_{Hu}^{a 3} + 2 q_{Hd}^{a 3}) + \sum_{r=1}^n q_S^{r 3} = 0. \label{N2}
\eea
Further linear constraints come from the requirement (\ref{yuk}) that all the Yukawa couplings be $U(1)'$ invariant. We will take (\ref{yuk}) to represent such linear system.

To begin with, we can make the following general statement: It is impossible to fulfill the chirality conditions (\ref{nomu}) with only one or two Higgs pairs, i.e. we must take $m=3$ in the minimal case. (In non-SUSY models, an attempt of assigning a different Higgs field to each family can be found in~\cite{Porto:2007ed}.)

This fact is true regardless of the number $n$ of singlets. It is easy to see that just one single pair ($H_u, H_d$) will not work by adding all the Yukawa conditions for the quarks
\be
     \forall i,~j: \quad q_Q^i + q_u^j +  q_{Hu} = q_Q^i + q_d^j +  q_{Hd} = 0.
\ee
Comparing with (\ref{L1}) we see that $q_{Hu} + q_{Hd} = 0$ violating (\ref{nomu}).
The same conclusions can be reached for two Higgs pairs ($m=2$). One has to consider all possible independent positions of the Higgs fields in the Yukawa terms and use only the anomaly conditions that do not involve the singlets. (The linear conditions (\ref{L1}), (\ref{L2}) and (\ref{L3}) are enough.)

With $m=3$ one can satisfy (\ref{nomu}), (\ref{noBL}) and (\ref{yuk}), together with (\ref{L1}), (\ref{L2}), (\ref{L3}) and (\ref{N1}) in many ways corresponding to different distributions of the Higgs fields into the Yukawa couplings. We chose the only combination that is phenomenologically allowed, namely the one that gives to the matter weak doublets $Q$ and $L$ family independent $U(1)'$ charges (see discussion in~\cite{ArkaniHamed:1999yy,Langacker:2008yv}). Thus, from now on, the Yukawa matter couplings in (\ref{yuk}) are taken to be
\be
   W \supset y^{u}_{ij} Q^j u^i H_u^i + y^{d}_{ij} Q^j d^i H_d^i + y^{e}_{ij} L^j e^i H_d^i, \label{ouryukawa}
\ee
where we use the same index for the Higgs fields since they are now associated to the family.

We can now start adding singlets. The systematic approach is to add one singlet at the time trying to preserve the above conditions.
An analysis similar to the one discussed above shows that it is impossible to fulfill the chirality condition (\ref{nomu}) with less than three singlets.
We thus assume $n=3$ singlets as well and search for all possible ways to couple them to the Higgs doublets in the superpotential.

In performing this search there is a further requirement coming from imposing the absence of unacceptably light charged Higgses.
This can be seen by setting the D-terms to zero and looking only at the F-terms and soft terms in the Higgs potential. This reduced potential must be invariant only under one global $SU(2)_L$ rotating all the Higgs doublets. If the non-abelian symmetry is larger, there are unwanted charged Goldstone modes whose mass cannot be lifted to an acceptable value by the D-terms~\footnote{The superpotential presented in the first version of this work suffered of this problem. It also had a non-minimal number of singlets (four).}.
Diagrammatically, denoting by a link $H_u^i \leftrightarrow H_d^j$ a term in the superpotential $S H_u^i H_d^j$ for some singlet superfield $S$, the graph obtained must be connected. This can be accomplished with a minimum of five such terms in the superpotential, but some of the singlets can be used in more than one coupling.

Taking into account all these requirements one can perform a computerized search for all models. In this letter we present a specific model that stand out for being characterized by a unique solution for the charges. This peculiarity arises from the fact that the cubic anomaly equation factorizes after having enforced all other constraints. The superpotential for this model read
\be
   W \supset \lambda_1 S^1 H_u^2 H_d^3 +  \lambda_2 S^1 H_u^3 H_d^2 + \lambda_3 S^3 H_u^1 H_d^2 + \lambda_4 S^3 H_u^2 H_d^1 + \lambda_5 S^2 H_u^3 H_d^3 +
   \frac{k}{2} S^2 S^3 S^3
   \label{W}
\ee
and the charges are given in table~1.

\begin{table}
  \centering
\begin{center}
\begin{tabular}{c|ccccccc}
 Family   & $Q$   & $u$       & $d$      &   $L$   & $e$      & $H_u$      & $H_d$ \\
\hline
  $i=1$   & $0$   & $3/10$    & $1/10$   &   $0$   & $1/10$   & $-3/10$    & $-1/10$ \\
  $i=2$   & $0$   & $-3/5$    & $-4/5$   &   $0$   & $-4/5$   & $3/5$      & $4/5$ \\
  $i=3$   & $0$   & $3/5$     & $2/5$    &   $0$   & $2/5$    & $-3/5$     & $-2/5$
\end{tabular}
\begin{tabular}{c|ccc}
Singlets & $S^1$   & $S^2$  & $S^3$  \\
\hline
         & $-1/5$  & $1$    & $-1/2$
\end{tabular}
\end{center}
\caption{\small \it Values of the $U(1)'$  charges $q_\psi^i$ for the model discussed in the text.}
  \label{chm1}
\end{table}

\section{Analysis of the model}

We now discuss some features of the model presented in the previous section. We begin by analyzing the scalar potential and discussing the tree level spectrum in the Higgs sector. We then look at the fermionic spectrum and finish with some comments on the $Z'$ mass and FV processes.

The only dimensionfull terms that are allowed in the full Higgs potential $V_H$ are the non-holomorphic, SUSY breaking diagonal scalar masses and the ``A-terms" given by the same analytic expression as the superpotential (\ref{W}) with the replacement of the coupling constants $\lambda_r \to b_r$ and $k \to a$. Being part of the soft lagrangian they can be naturally at the TeV scale if generated via dynamical SUSY breaking. In this case however the extra $U(1)'$ is not enough to prevent another source of LFV such as $m_{L ij}^2 L^{i\dagger} L^j,$ but we will assume that the soft terms are generated in a gauge mediation framework, where such problems do not arise\footnote{This is not strictly true in our context, since, as we will discuss in section~4, the $Z'$ couplings do violate flavor symmetry and thus will generate subleading off diagonal terms.}. The scalars in the matter sector can always be arranged to be non-tachyonic by suitably large soft masses. We thus set them to zero and analyze the scalar potential $V_H$ in the Higgs sector alone. We also assume the couplings to be real throughout the analysis.

The model (\ref{W}) has one additional global Peccei-Quinn (PQ) symmetry giving rise to an axions in the neutral CP-odd sector. (This is actually a generic feature of all models of this type.) We assume that its coupling is suppressed by a DFSZ mechanism \cite{Zhitnitsky:1980tq, Dine:1981rt} at some higher scale. Furthermore, the SUSY part of the potential has flat directions that could be lifted by higher dimensional operators such as $S_2 S_1^5$, also suppressed by some higher scale. Once all flat directions are lifted, the addition of soft terms does not reintroduce instability at large field since they are at most cubic in the fields. In addition, these higher dimensional terms break explicitly the PQ symmetries giving an extra contribution to the mass of the axions. We will not study the effect of such terms in detail and instead focus in the renormalizable part of $V_H$ allowing for possible metastability.

\begin{figure}
\centering
\includegraphics[width=.45\textwidth,clip]{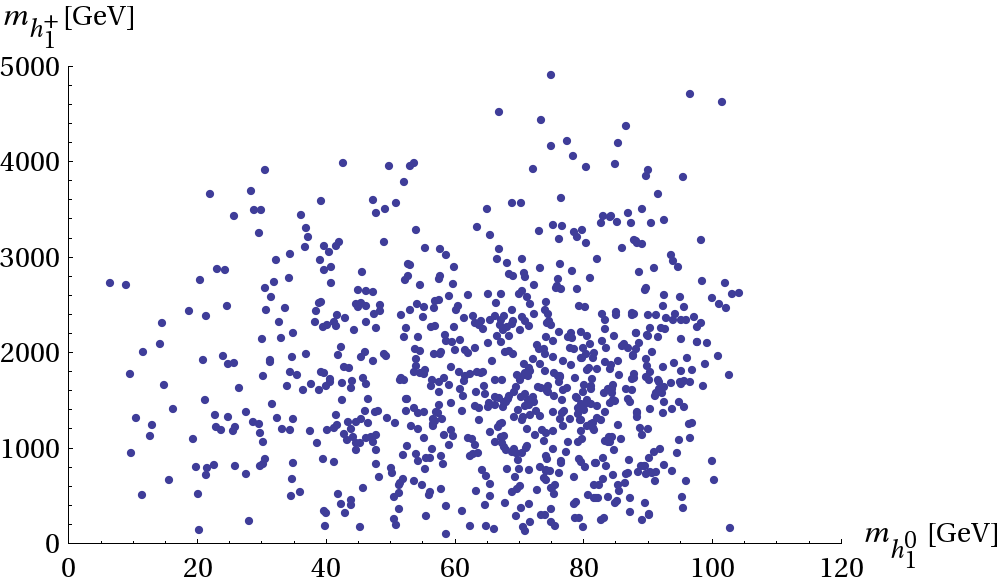}
\hfill
\includegraphics[width=.45\textwidth,clip]{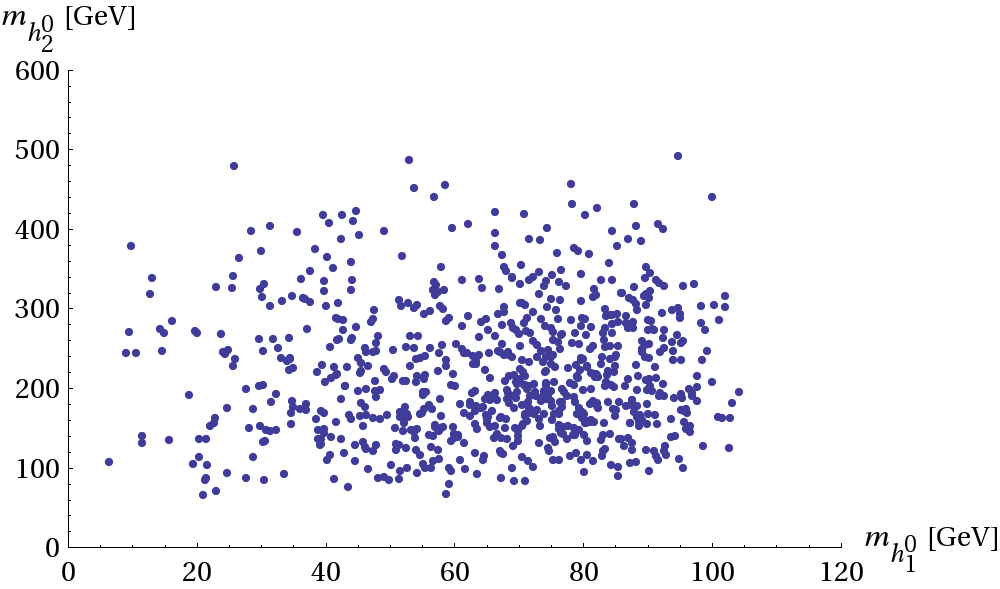}
\caption{\small \it On the left: Tree level mass of the lightest CP even Higgs vs. that of the lightest charged Higgs for a set of randomly generated values of the couplings. On the right: Tree level mass of the lightest CP even Higgs vs. that of the next to lightest one for the same set of couplings.}
\label{plHiggs}
\end{figure}

Not counting the would-be Goldstone bosons and the axion, the model (\ref{W}) has 5 charged, 9 CP-even and 6 CP-odd massive Higgses.
We perform a generic random scan of parameter space. As the only simplifying assumption we set the vev's of the neutral parts of the first two pairs of Higgs doublets to $v < 12$~GeV , while letting $v_u^3$ and $v_d^3$ vary in the $100$~GeV range, subjected, of course, to the constraint on the $Z$ mass. We take the vevs of the singlets to vary between $3$ and $6$~TeV and all the coupling constants in the range ${[-1, 1]}$. We then solve for the soft masses by imposing the extremality of the potential and look for islands of local stability. The scatter-plots in fig.~\ref{plHiggs} indicate the tree level masses for the lightest scalars obtained this way. These regions occur for $0.2< |k| < 0.5$ and $a < -0.2$. Note that the tree level mass of the lightest Higgs can be pushed over the $Z$ mass, although there are also points where it is unacceptably light.

\begin{figure}
\centering
\includegraphics[width=.45\textwidth,clip]{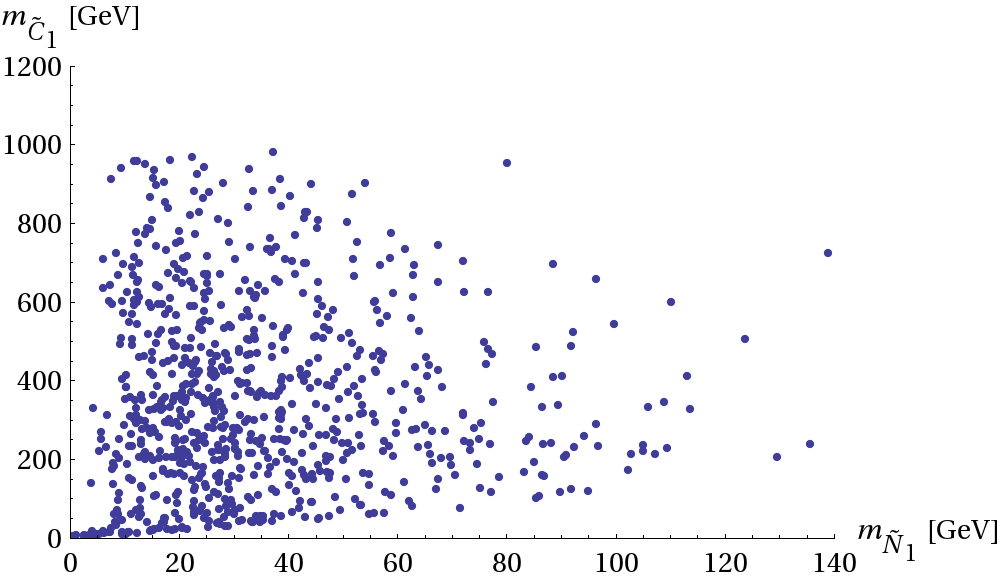}
\hfill
\includegraphics[width=.45\textwidth,clip]{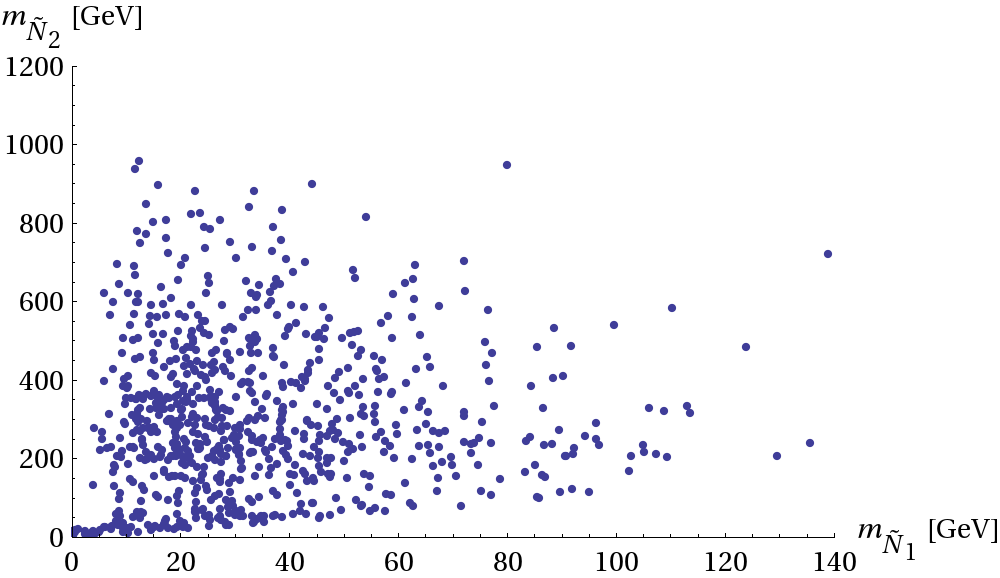}
\caption{\small \it  On the left: Tree level mass of the lightest neutralino vs. that of the lightest chargino for a set of randomly generated values of the couplings, including the gaugino Majorana masses. On the right: Tree level mass of the lightest neutralino vs. that of the next to lightest one for the same set of couplings.}
\label{plDM}
\end{figure}

The model also gives rise to 4 charginos and 12 neutralinos and the scatter-plot of the lightest chargino vs. the lightest neutralino and of the two lightest neutralinos are shown in fig.~\ref{plDM}, where we have also randomized over the gaugino Majorana masses $M_1, M_2, M'$ in the range between $0.2$ and $1$~TeV.
The generic feature of the model is a very light neutralino LSP. For 89\% of the points examined we also find that the next to lightest particle in this sector is also a neutralino, the lightest chargino being typically the third one.

In this context, it is worth mentioning that the lightest CP even scalar ($h^0_1$) consists mostly of the neutral Higgs fields of the third generation $H_u^{0\,3}$ and $H_d^{0\,3}$ (in case of tree level mass being around $100$~GeV it also contains some contribution from the singlets). The lightest neutralino is to high accuracy the fermionic component of the singlet $S^1$, alleviating the problems associated with invisible decays. (Recall that the singlets only couple to the Higges and the $Z'$.) The higher mass particles are typically in the multi-TeV range.

Finally, the mass matrix for the vector bosons depends only on three combinations of the Higgses vevs, which are, now for generic values of the vevs:
\bea
    v_1^2 &=& \sum_i \left(|v_u^i|^2 + |v_d^i|^2\right)\approx 174~{\mathrm{ GeV}}^2, \quad  v_2^2 = \sum_i \left(q_{Hu}^i |v_u^i|^2 - q_{Hd}^i |v_d^i|^2\right), \nn\\
          & & v_3^2 = \sum_i \left((q_{Hu}^i)^2 |v_u^i|^2 + (q_{Hd}^i)^2 |v_d^i|^2 + \sum_r (q_S^r)^2 |v_S^r|^2 \right).
\eea
Notice that $v_2 < v_1 \ll v_3$ because of cancelations in $v_2$ and the presence of singlet vevs in $v_3$.

The masses of the charged vector bosons are unaffected while the masses of the two neutral massive bosons are, to first non-trivial order in $1/v_3^2$:
\be
    m^2_Z \approx \frac{1}{2} (g_1^2 + g_2^2) v_1^2 - \frac{1}{2} (g_1^2 + g_2^2) \frac{v_2^4}{v_3^2}, \qquad
    m^2_{Z'} \approx 2 g'^2 v_3^2 + \frac{1}{2}(g_1^2 + g_2^2)\frac{v_2^4}{v_3^2}.
\ee
Recent experimental bounds on $ m_{Z'} $ can be found in~\cite{Collaboration:2011dca,Chatrchyan:2011wq}.
The presence of a $Z'$ comports a shift upwards of the $\rho$ parameter of the order of $v_2^4/v_3^2 v_1^2$ that must be $< {\mathcal{O}}(10^{-3})$ in order to satisfy precision tests. To make a complete analysis however, one should also include the loop contributions of the four extra Higgs doublets on top of the MSSM one~\cite{Drees:1990dx}, but this is beyond the scope of the present letter.

The strongest experimental constraints on this model arise from the potential FV effects. These are induced by two sources, namely the family dependent charges and the multiple Higgs couplings.

Tree level FV from the gauge sector arises after rotating the (right handed) matter fermions into their mass eigenstates. Their effects on the low energy physics have been analyzed in e.g.~\cite{Langacker:2000ju} and can be parameterized by the off diagonal elements of the matrix
\be
         B^\psi_R = V_R^\psi \; q_R^\psi\; V_R^{\psi\dagger},
   \label{Bs}
\ee
where we denote by $V_L^\psi$ and $V_R^\psi$ ($\psi = u, d, e$) the unitary matrices that diagonalize the Yukawa couplings after EW breaking and by $q_R^\psi$ the diagonal matrix of charges.

In the ``best case" scenario, where the diagonalization of the Yukawa couplings is achieved almost entirely by $V_L^\psi$ and $V_R^\psi\approx \mathbf{1}$, the matrices  $B^\psi_R$ are also close to be diagonal.

Bounds on the off-diagonal elements of $B^\psi_R$ come from various FV processes such as meson mass splitting~\cite{Nakamura:2010zzi},
$\mu^- \to e^- \gamma$ decay~\cite{Adam:2011ch} and muon conversion~\cite{Wintz:1998rp,Suzuki:1987jf, Bernabeu:1993ta, Kuno:1999jp}.
For our model, with $m_{Z'}$ in the TeV range, all this can be essentially summarized by saying that, for $i\neq j$, $(B^\psi_R)_{ij} \lesssim 10^{-5}$, requiring $V_R^\psi\approx \mathbf{1}$ to the same accuracy.

The presence of multiple Higgs doublets implies additional sources of FV from the Yukawa couplings. (A similar example arises in the type III 2HDM discussed e.g. in~\cite{Branco:2011iw}.) However, because of the structure of the Yukawa couplings of our models (\ref{ouryukawa}), the FV processes are controlled by the same matrix $V_R^\psi$, since for $V_R^\psi = \mathbf{1}$ there is a basis in which their mass matrix receives contribution from a single source. The bounds on the off diagonal terms are similar to the previous ones.

\subsection*{Conclusions}

In this letter we presented an extension of the MSSM where the absence of $\mu$-terms and R-parity violating interactions are justified by an extra $U(1)'$ symmetry. This was achieved by extending the Higgs sector and giving family dependent charges to the right-handed matter fields.

It could be worthwhile to study models of such type in more detail. We presented the main theoretical ideas of this construction and the basic phenomenological features. From the preliminary analysis of the spectrum, we see that the most striking characteristics are the fact that the tree level Higgs mass can be raised above the $Z$ mass and the presence of a light neutralino, which is mostly a singlet. Most of the remaining scalars and neutralinos/charginos end up in the multi-TeV range, so we expect that the LHC signatures will not be too different from those of some corners of the MSSM parameter space, and even more closely resemble those of the NMSSM. The large number of additional parameters makes a more detailed analysis of models of this type more complicated. This analysis would certainly become much more motivated in the presence of a hint of SUSY or extra $Z$ bosons at the LHC.

\subsection*{Acknowledgments}

We thank R. Argurio, M. Bertolini, A. Mariotti, C. Petersson and A. Romanino for reading and commenting on an early draft of the manuscript. We thank in particular A. Romanino for many enlightening email exchanges and for help with references.
We would also like to thank C. Forss\'{e}n, T. Nilsson, J. Olsson, R. Pasechnik, J. Rathsman and P. Salomonson for useful discussions.

The research of G.F. is supported in part by the Swedish Research Council (Vetenskapsr{\aa}det) contract B0508101.

\end{document}